\documentclass[useAMS,usenatbib]{mnras}
\usepackage{graphicx}
\graphicspath{{images_de_larticle/}}
\usepackage{amsmath}

\usepackage[usenames,dvipsnames]{color}

\def\lya{Ly$\alpha$~}

\title[The timing of the percolation of HII bubbles at the end of reionization with  ULAS J1120+0641]{
A tale of seven narrow spikes  and a long trough: constraining the  timing of the percolation of HII bubbles at the tail-end of reionization with  ULAS J1120+0641.}
\author[J. Chardin et al.]{Jonathan Chardin$^{1}$,$^{2}$\thanks{E-mail:jc@ast.cam.ac.uk} , Martin G. Haehnelt$^{1}$, Sarah E.I. Bosman$^{1}$
and Ewald Puchwein$^{1}$\\
$^{1}$Kavli Institute for Cosmology and Institute of Astronomy, Madingley Road, Cambridge CB3 0HA\\
$^{2}$Universit\'e de Strasbourg, CNRS, Observatoire Astronomique de Strasbourg, UMR 7550, F-67000 Strasbourg, France\\}

\begin{document}


\date{Accepted / Received }

\pagerange{\pageref{firstpage}--\pageref{lastpage}} \pubyear{2015}

\maketitle

\begin{abstract}
High signal-to-noise observations of the \lya forest transmissivity in the  $z=7.085$ QSO ULAS J1120+0641 show seven narrow transmission spikes 
followed by a long  240 cMpc/h trough. Here we use radiative transfer simulations of cosmic reionization previously calibrated to match a wider range 
of  \lya forest data to show  that the occurrence of  seven transmission spikes in the narrow redshift range $z=5.85-6.1$ is very sensitive 
to the exact timing of reionization. Occurrence of the spikes requires the most under dense regions of 
the IGM  to be already fully ionised.  The rapid onset of a long trough at $z=6.12$ requires a strong 
decrease of the photo-ionisation rate $\Gamma$ at $z\ga 6.1$ in this line-of-sight,  consistent with the end of percolation at this redshift. 
The narrow range of reionisation histories  that we previously found to be consistent with a wider range of \lya forest data 
have a  reasonable probability of showing seven  spikes and the  mock absorption spectra provide an excellent match to the spikes and the trough in the observed spectrum of ULAS J1120+0641. Larger samples of high signal-to-noise searches for rare \lya transmission 
spikes at $z>5.8$ should therefore provide important further  insights into the  exact timing  of the percolation of HII bubbles at the tail-end of 
reionization. 
\end{abstract}



\section{Introduction}
\label{intro}

The Lyman alpha (Ly$\alpha$) forest is the main probe of the physical state of the intergalactic medium (IGM) after hydrogen reionization 
(See \citealt{2015PASA...32...45B} and \citealt{2016ARA&A..54..313M} for recent reviews).
The rapid increase of the \lya optical depth at $z>5$ suggests that at $z\sim6$ we are seeing the last stages of the hydrogen reionization process
(\citealt{2006AJ....132..117F}; \citealt{2013MNRAS.430.2067B}; \citealt{2015MNRAS.447.3402B}).
The rapid decrease of the \lya emitters space density at $z>7$ (\citealt{2013MNRAS.429.1695B} and \citealt{2014MNRAS.440.3309D}) 
and the low value of the Thomson optical depth suggested by recent Planck data 
(\citealt{2015arXiv150201589P} and \citealt{2016arXiv160503507P}) would indicate  a somewhat smaller duration 
and a later start of the reionization process than  suggested by the WMAP results.

Currently \lya forest studies allow us to probe the redshift evolution of the ionisation state of the IGM  up to redshift $z\sim6$.
However, because the scattering cross-section of the \lya transition is very high, it becomes  rather  challenging to use the forest to put constraints on 
the mean neutral hydrogen fraction $x_{\mathrm{HI}}$ at $z\ga 5.7$. When  the neutral hydrogen fraction at mean density  exceeds values of $x_{\mathrm{HI}}\ga 10^{-4}$ at $z\sim 6$, this  results  in  average \lya optical depths of a few and an almost completely absorbed 
\lya forest. Moreover, the apparent decline of  the space density of bright QSOs  as we enter the reionization period  allows us to  probe the evolution of the reionization process only with rather small samples with only a handful of  objects detected at $z>6.5$  (\citealt{2011Natur.474..616M}, \citealt{2013ApJ...779...24V}, \citealt{2015ApJ...801L..11V}, \citealt{2016ApJ...828...26M}, \citealt{2017arXiv170405854M}, \citealt{2017MNRAS.468.4702R} and \citealt{2017MNRAS.466.4568T}).

Recently, \citet{2017A&A...601A..16B} (B17 hereafter) obtained a very high S/N spectrum of the QSO with the highest redshift known,   
ULAS J1120+0641 at z=7.084 (\citealt{2011Natur.474..616M}). B17  used  \textit{Hubble Space Telescope (HST)} imaging to flux-calibrate 
and combined additional  data  with their previous spectroscopic observations with the X-shooter instrument at the Very Large Telescope (VLT).
They reported the detection of seven narrow transmission spikes in the redshift range  $5.858<z<6.122$ followed  
by the longest  Gunn-Peterson trough (\citealt{1965ApJ...142.1633G}) of complete absorption known in a QSO  spectrum.
The long  trough extends from $z=6.122$ up to $z=7.04$ at the edge of the QSO  near zone and corresponds to a length 
of 240 cMpc/$h$.

B17 then used the Sherwood simulation (\citealt{2017MNRAS.464..897B}) to constrain the redshift evolution of 
the \lya  optical depth  at $z>5.7$  by trying to reproduce the 240 cMpc/$h$ long absorption 
trough in mock absorption spectra produced from the simulation. They concluded that the occurrence of this  long 
GP trough provides only weak additional constraints on the evolution of the neutral hydrogen fraction ($\mathrm{x_{HI}\ge10^{-4}}$)
much weaker than the constraints inferred from the red damping wing by  \citealt{2011Natur.474..616M} ($\mathrm{x_{HI}>0.1}$), 
but see also Bolton et al (2011) and Bosman \& Becker (2015).
B17 used their analysis to provide an   updated  power-law parametrization of the effective optical depth  $\tau_{\mathrm{eff}}(z)$ 
constrained by the overall transmitted flux in the redshift range  $5.858<z<6.122$.   They however did not try to 
reproduce the observed transmission spikes in detail.

Here we  will  model both the observed transmission spikes and the long trough 
in the spectrum of ULAS J1120+0641 with radiative transfer simulations that 
were originally introduced in \citealt{2015MNRAS.453.2943C}). 
Our main goal is to model  the transmission spikes identified by B17  
in the observed spectrum of ULAS J1120+0641 (see also \citealt{2006MNRAS.370.1401G} for some modelling of the dependence of high-redshift 
transmission spikes  on the reionization history based on a semi-analytical model  of QSO absorption spectra). 
In \citealt{2015MNRAS.453.2943C}, we have presented a set of radiative transfer simulations carefully calibrated with  \lya forest data.
We have shown that these simulations reproduce the observed evolution of 
$\tau_{\mathrm{eff}}(z)$ and the inferred hydrogen photo-ionisation rate $\Gamma$ very well. 
These simulations  also  reproduce the evolution of the average mean 
free path $\mathrm{\lambda_{mfp}^{912}}$ as measured by \citet{2014MNRAS.445.1745W} in the post reionization period and are  consistent with the latest measurement of the  Thompson optical depth inferred by Planck (\citealt{2016arXiv160503507P}). The simulations  follow self-consistently the growth and percolation of HII bubbles and are  well suited to put constraints on the exact timing of the end of reionization
by comparing the occurence of transmission spikes at the tail-end of reionization in  simulated spectra with that in observed QSO absorption spectra.  

The paper is organized as follows.
In Sect. \ref{Simu}, we introduce our numerical simulations and the different steps to construct \lya spectra that mimic the observed spectrum of ULAS J1120+0641.
Sect. \ref{results} presents our analysis of the mock absorption spectra  and the comparison with the observed spectrum.
Finally in Sect. \ref{discussion} we discuss our results and their implications for future observations to 
constrain in more details the evolution of the ionization history at the tail-end of reionization before giving our conclusions in Sect. \ref{conclusion}.
Throughout the paper, we use the following cosmology (as derived from the 2014 Planck temperature power spectrum data alone, \citealt{2014A&A...571A..16P}) : 
$\Omega_m = 0.3175$, $\Omega_{\Lambda} = 0.6825$,
$\Omega_b = 0.048$, $h = 0.6711$, $\sigma_8 = 0.8344$, and $n_s = 0.9624$.

\section{Methodology}
\label{Simu}

Here we briefly describe the  radiative transfer simulations and our methodology to construct mock \lya forest spectra that mimic the 
observed spectrum of ULAS J1120+0641.

\begin{figure}
   \begin{center}
      \includegraphics[width=\columnwidth]{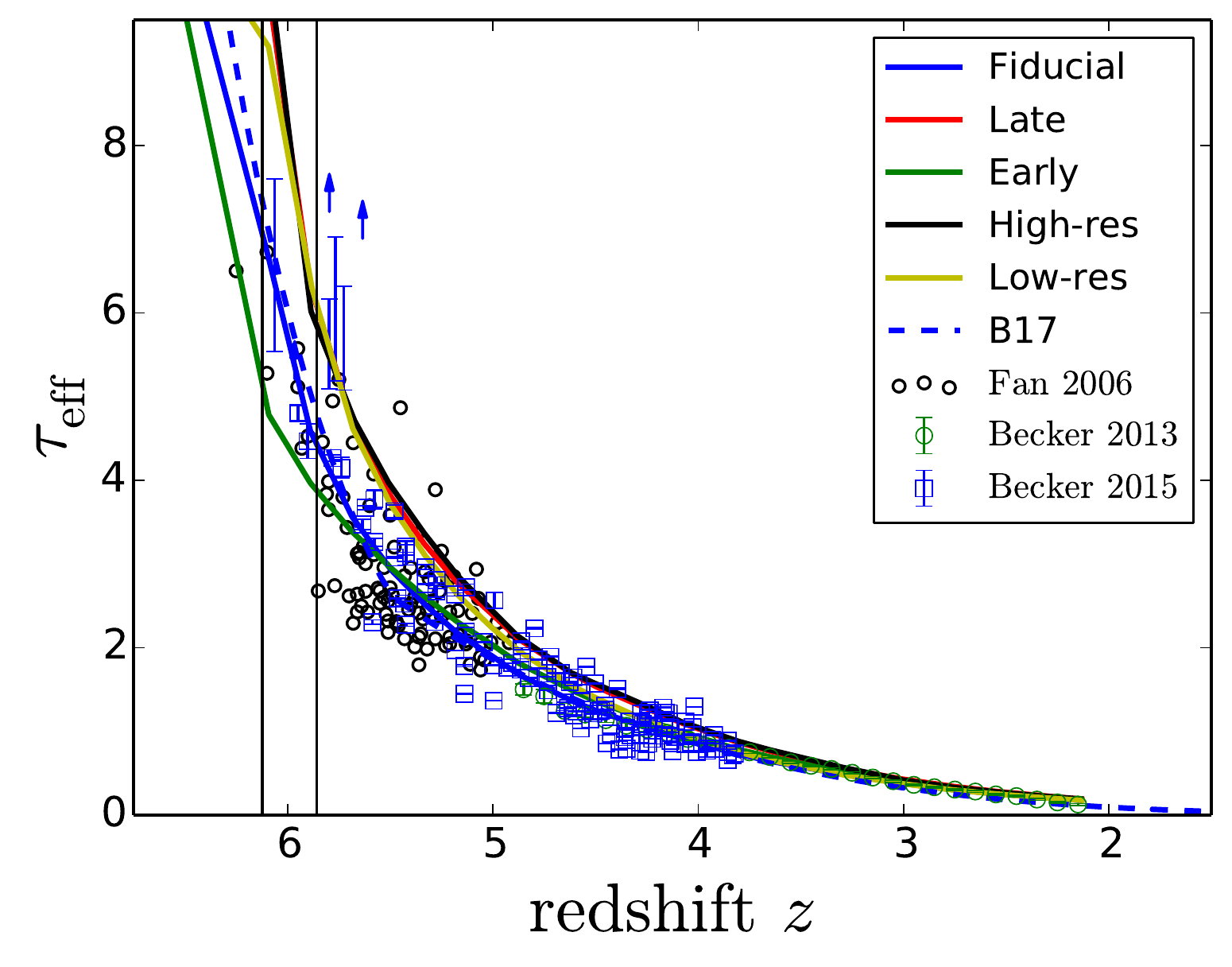}   
  \caption{Evolution of the effective Lyman alpha optical depth $\tau_{\mathrm{eff}(z)}$ in our different models.
The two black vertical lines show the redshift range where the seven spikes have been detected in ULAS J1120+0641. 
The blue dashed line show the evolution from B17 which matches the evolution in our  fiducial model very well.}
    \label{teff}
  \end{center}
 \end{figure}

\begin{figure}
   \begin{center}
      \includegraphics[width=\columnwidth]{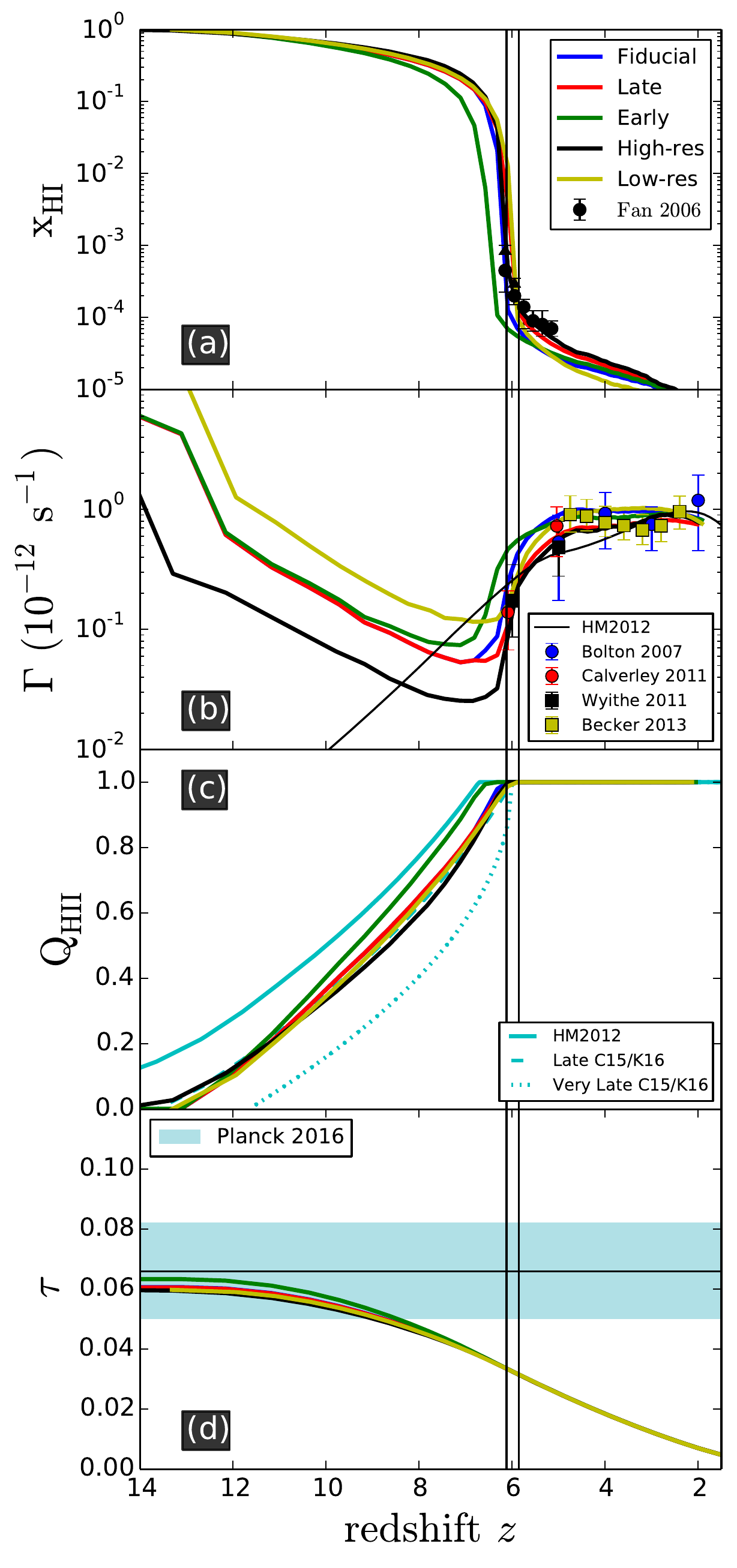}   
  \caption{ Panel (a) : redshift evolution of the average neutral fraction in our different models.
Panel (b) : redshift evolution of the average photoionization rate $\left<\Gamma_{12}=\Gamma/10^{12}\right>$.
Panel (c) : redshift evolution of the volume filling factor $\mathrm{Q_{HII}}$ of HII regions.
The three different cyans lines show the evolution of $\mathrm{Q_{HII}}$ in the three models presented in \citet{2015MNRAS.452..261C} and \citet{2016MNRAS.463.2583K}.
Panel (d) : redshift evolution of the Thompson optical depth.}
    \label{reion_history}
  \end{center}
 \end{figure}

\subsection{The radiative transfer simulations} 
\label{simulation}

Our reionization simulations are described in more detail in our  recent studies (\citealt{2015MNRAS.453.2943C}; \citealt{2017MNRAS.465.3429C})  and have been performed  in two steps. 
First hydrodynamic simulations were run and  the radiative transfer calculations were then  performed  by post-processing  the hydrodynamic simulations for a range  of ionizing source models.

In our simulations the evolution of dark matter and the hydrodynamics of the gas 
were performed  with the RAMSES code (\citealt{2002A&A...385..337T}).  
All the simulations presented in this work use a coarse grid with $512^3$ cells.
The UV background of \citet{2012ApJ...746..125H} was used in the RAMSES simulations to model the  evolution of the thermal and ionisation state 
of the IGM in the hydrodynamic simulation   which  results in good agreement with observations from the \lya forest in the post-reionization Universe
(\citealt{2015MNRAS.450.4081P}). As discussed in detail in \citet{2015MNRAS.453.2943C} the temperature  of the hydro-simulation were then also used in the following 
radiative transfer calculations that were performed in  post-processing with the ATON code (\citealt{2008MNRAS.387..295A}).

ATON is a radiative transfer code utilising a moment based description of the radiative transfer equation ported on GPU architecture to accelerate the calculations.
We assumed all ionizing photons to have  an energy of 20.27 eV in all simulations.
Because of our monofrequency treatment high energy ionizing photons are not modelled properly. 
Therefore we use the temperature from the RAMSES simulation to be consistent with the 
observed evolution in the post-reionization IGM. Note that we will later also discuss simulations where we have changed this temperature.
Radiative transfer ATON outputs  were created  for snapshots of the RAMSES  simulations separated by 40 Myrs. 

Our ionizing source modelling is assuming the dark matter haloes identified in the RAMSES simulations as the sources of ionizing photons and they are assumed to emit continuously.
The ionizing luminosity is assumed to scale with the mass of dark matter haloes in a similar way as in \citet{2006MNRAS.369.1625I} 
(but see also \citealt{2012A&A...548A...9C} and \citealt{2014A&A...568A..52C}).
The normalisation  varies with redshift in such a way that the integrated  comoving ionizing emissivity is 
similar to that of the \citet{2012ApJ...746..125H} UV background model, but somewhat modified in order to get a better match with the 
observations of the hydrogen photo-ionisation rates.
In \citet{2015MNRAS.453.2943C} we  have shown that our  calibrations of the ionising emissivity obtained in this way 
are also consistent with the redshift evolution of the luminosity function of high redshift galaxies. 

\renewcommand{\arraystretch}{1.5}
\setlength{\tabcolsep}{0.5cm}
\begin{table}
\begin{center}
\begin{tabular}{|c|c|c|c|c|c|c|}
  \hline
  Model name & Box size &Particle number \\
  \hline
  512-20-good\_h (Fiducial) & 20 & $512^3$  \\
 \hline
  512-20-good\_l (Late) & 20 & $512^3$ \\
\hline
  512-20-early (Early) & 20 & $512^3$  \\
\hline 
   512-40 (Low-res) & 40 & $512^3$  \\
\hline 
   512-10 (High-res) & 10 & $512^3$  \\
\hline  
\end{tabular}
\caption{Summary of the different simulations studied in this work.}
\label{tab1}
\end{center}
\end{table}

\subsection{Set of models} 
\label{set_model}

We will study here different models for the reionization history.
All these different simulations have already been discussed in \citet{2015MNRAS.453.2943C}.
Table \ref{tab1} summarizes some of the properties of the different models.

Our reference model is the 512-20-good\_h  model of  \citet{2015MNRAS.453.2943C}
 where 512 stands for the number of cells in one dimension  and 20 for  the box size in cMpc/$h$.
We also consider two others models here with a slightly later/earlier reionization history compared to the reference model.
The model were reionization occurs slightly later is the 512-20-good\_l and the  model where it occurs slightly earlier 
is the 512-20-early model from \citet{2015MNRAS.453.2943C}.

Finally, we also consider  two further simulations  to  study the impact of the resolution of the simulations.
For these we used the models 512-40 and 512-10 of late reionization also described in \citet{2015MNRAS.453.2943C} with 512$^3$ grid cells  as in the others models, but with a box size of 40 and  10 cMpc/$h$, respectively. This  allows us to test the effect of raising/lowering the mass resolution by 
a factor  eight compared to our fiducial model with a box size of 20 cMpc/$h$.

Fig. \ref{teff} shows the evolution of the effective optical depth $\mathrm{\tau_{eff}}$ in our different simulations.
The blue dashed line shows the evolution inferred by B17, parametrized as

\begin{equation}
 \tau_{\mathrm{eff}}(z)
\left\lbrace
\begin{array}{ccc}
0.85\left(\frac{1+z}{5}\right)^{4.3}  & \mbox{for} & z\le5.5,\\
 \\
2.63\left(\frac{1+z}{6.5}\right)^{11.2} & \mbox{for} & z>5.5.
\end{array}\right.
\label{teff_B17}
\end{equation}

Interestingly, our  fiducial model matches  almost exactly the evolution  of   $\mathrm{\tau_{eff}}$ inferred by  B17.
We therefore expect our mock absorption spectra constructed for this model to be in good  agreement with the observed spectrum.

In Fig. \ref{reion_history}, we show the redshift evolution of a number of volume averaged  quantities in our different models such as
the mean neutral hydrogen fraction, the mean photoionization rate $\mathrm{\Gamma}$ and
the Thompson optical depth.   We can see that our simulations bracket well a  range of  observational constraints. 
In panel (c) we compare the redshift evolution of the volume filling factor of ionised region for our five different 
reionization histories with those presented in \citet{2015MNRAS.452..261C} and \citet{2016MNRAS.463.2583K}.
Note that all reionization histories are rather similar to the  `late' model in \citet{2015MNRAS.452..261C} and \citet{2016MNRAS.463.2583K}
that these authors have shown also to be consistent with the rapid evolution of \lya emitters at $z>5.5$.

We will  here use  the occurrence of transmission spikes followed by a  long fully absorbed GP trough starting at $z\approx 6.12$ to  constrain the timing of 
reionization in the line-of-sight to ULAS J1120+0641.

\begin{figure*}
   \begin{center}
      \includegraphics[width=\textwidth,height=\textheight,keepaspectratio]{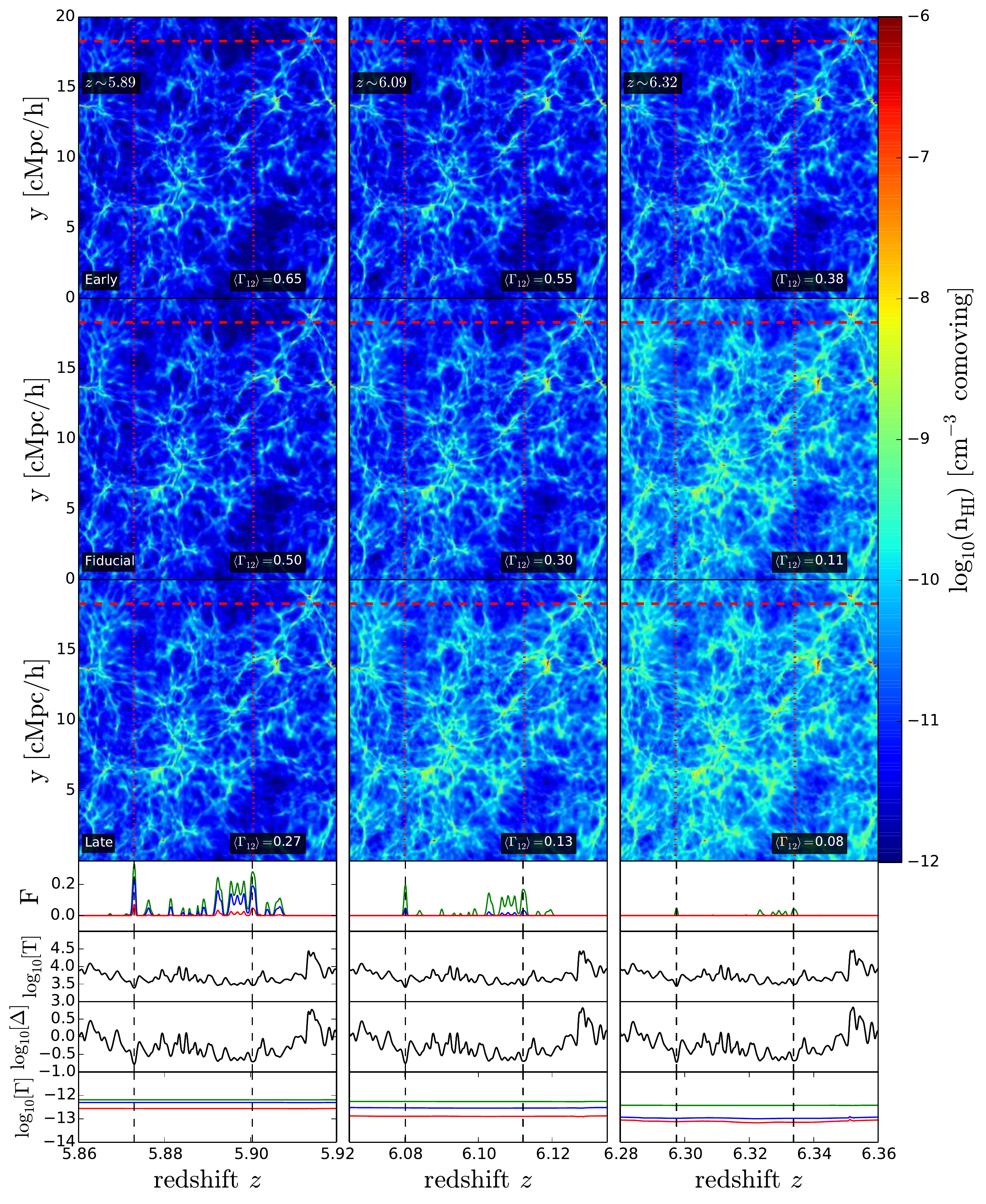}   
  \caption{Example of a line of sight (shown with the red dashed horizontal lines) through the volume of our three 20 cMpc/$h$ simulations at three different redshifts.
For each model we show the transmitted flux $\mathrm{F}$, as well as the temperature T,  the overdensity $\mathrm{\Delta}$ 
and the photoionization rate $\mathrm{\Gamma}$ in velocity space along the line-of-sight. The background shows the neutral hydrogen comoving density $\mathrm{n_{HI}}$
in the different models in a slice of 39.0625 kpc/$h$ thickness.}
    \label{los_plus_light_cone}
  \end{center}
 \end{figure*}

\begin{figure*}
   \begin{center}
      \includegraphics[width=\textwidth,height=\textheight,keepaspectratio]{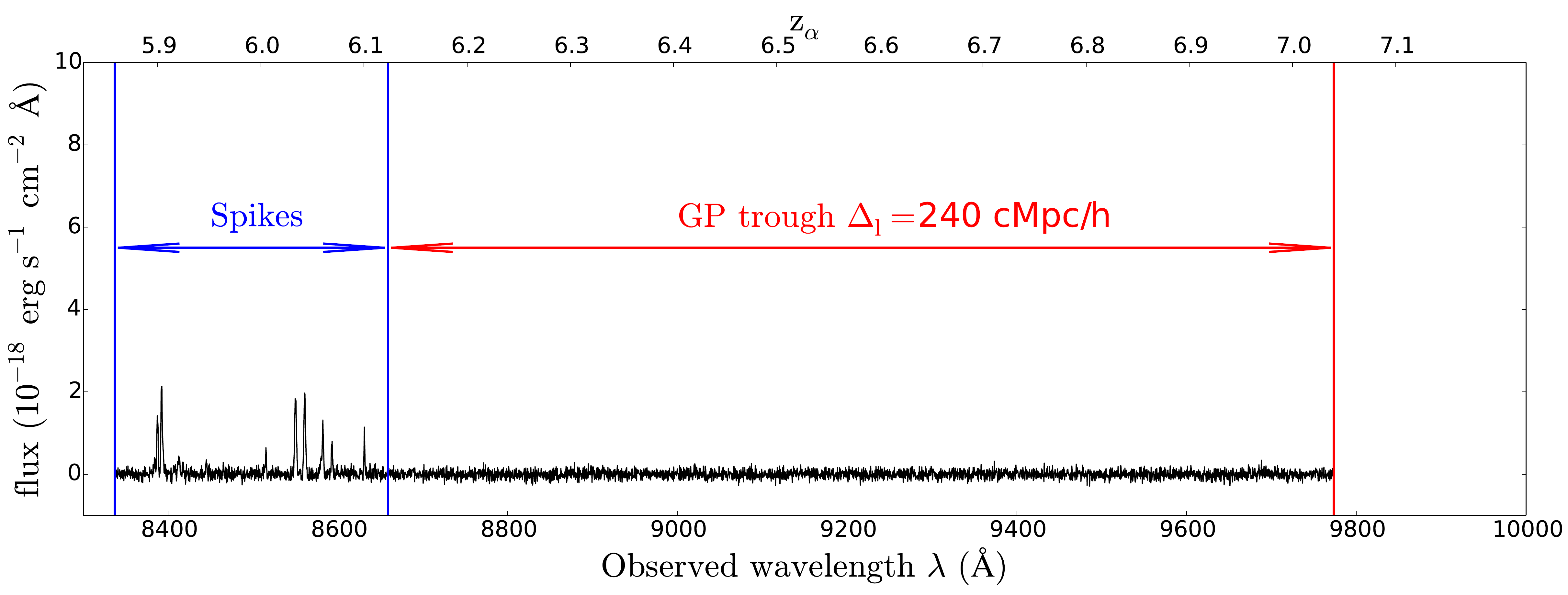}   
  \caption{Example of a synthetic Lyman alpha forest spectrum corresponding to the wavelength 
range of the observed  spectrum of ULAS J1120+0641 presented by B17.
The simulated spectrum has been first convolved to the instrumental resolution with a Gaussian of FWHM=34 km/s 
and then rebinned to a pixel size of 10 km/s as in the  data presented in  B17.
Random Gaussian noise has been  added to every pixel with a standard deviation corresponding 
to the average value of the observed spectrum presented in B17. 
The spectrum is divided into two sections that correspond to the wavelength range of the observed spikes in ULAS J1120+0641  
and to that of the long Gunn-Peterson trough of 240 cMpc/$h$ length.}
    \label{all_spectrum}
  \end{center}
 \end{figure*}

\subsection{Creating synthetic Lyman alpha forest spectra from the simulations} 
\label{mock_spectra}

For this study, we have generated mock \lya forest spectra that mimic the observed spectrum 
of ULAS J1120+0641 from the five different radiative transfer simulations of cosmic reionization described above.
As already discussed seven statistically significant transmission spikes have been reported for 
the redshift range $5.858<z<6.122$  by  B17.
This redshift interval corresponds to a comoving length of 112.9 cMpc for our cosmology, equivalent to  75.77 cMpc/$h$.
This is roughly  four times  our fiducial box size of 20 cMpc/$h$.
B17 also reported the longest fully absorbed  Gunn-Peterson trough ever found in a QSO  spectrum with a length  of 240 cMpc/$h$ 
from $z=6.122$ up to $z=7.04$ at the edge of the quasar near zone. For the slightly different 
cosmological  parameter  adopted in  our simulations, this  corresponds to a comoving length of 234 cMpc/$h$.
The  part of the spectrum of interest for our analysis has therefore a total length of $\sim$ 310 cMpc/$h$.

In our radiative transfer simulation with ATON, outputs were generated every 40 Myrs from redshift $z\sim100$ up to redshift $z\sim2$.
We use outputs that fall inside the redshift range corresponding to the location of the observed spikes and  trough to 
construct spectra that cover a total comoving length of $\sim310$ cMpc/$h$.
In practice, we concatenate random line-of-sights through our simulation outputs to construct the skewers.
For each simulation, we calculated 1000 such skewers that mimic the observed spectrum.

In Fig. \ref{los_plus_light_cone}, we show the normalized transmitted flux $\mathrm{F=e^{-\tau}}$  
along the same line of sight for the fiducial, late and early  reionization histories for our simulations  
with box size of 20 cMpc/$h$. The three redshift ranges covered by the skewers for which the spectra 
are shown are similar  to those where spikes occur  in the observed spectrum of ULAS J1120+0641.
We show also the underlying neutral hydrogen $\mathrm{n_{HI}}$ density field in the slice where the line of sight is calculated, 
the value of the photoionization rate $\mathrm{\Gamma}$ 
along the line-of-sight, as well as the over density and the temperature. For ease of comparison with the absorption spectrum 
the temperature, over-density and photoionization rate along the line-of-sight  are plotted in velocity rather than real space. 
We can see that the locations 
where  transmitted flux is reported are consistent from one model to another for the same line of sight.
However, the amount of transmitted flux is different in the different models. As the underlying density and temperature field is the same 
in the three models,  the difference in transmitted flux is here  only due to differences in the photoionization rate values along the skewer
due to  the different reionization histories.  The transmission spikes all coincide with the most under-dense regions in our simulations. 
For all three different reionization histories percolation is  already complete  and the spatial fluctuations in the photo-ionisation 
rate are small.

To construct mock spectra directly comparable to the observed spectrum, we transformed our optical depth in each pixel to a flux in $\mathrm{erg \,s^{-1} \, cm^{-2} \, \AA^{-1}}$.
We use the normalisation at 1280 $\mathrm{\AA}$ in the observed spectrum and assume a power law for the normalisation as a function of observed wavelength such that 
\begin{equation}
 f_{\lambda}=\left[\frac{\lambda}{1280\times(1+z)}\right]^{-0.5}f_{1280}
\end{equation}
with $f_{1280}=5.5\times10^{-18} \, \mathrm{erg \,s^{-1} \, cm^{-2} \, \AA^{-1}}$.
Then we tranform our normalized transmitted flux $\mathrm{F=e^{-\tau}}$ in an observed flux as $\mathrm{F_{\lambda}=f_{\lambda}\times F}$.
 
In order to mimic the observed spectrum, we smooth the mock spectra with a Gaussian kernel with FWHM=$34\, \mathrm{km\,s^{-1}}$.
We then rebin the spectrum with a pixel size of $10\, \mathrm{km\,s^{-1}}$ and add random Gaussian noise to the spectra with a dispersion given by the ULAS J1120+0641 error spectrum with 
$\sigma=0.125 \times 10^{-18} \, \mathrm{erg \,s^{-1} \, cm^{-2} \, \AA^{-1}}$.  

In Fig. \ref{all_spectrum}, we show one of our  mock spectra constructed from our fiducial simulation.  
Gratifyingly there are several statistically significant transmission spikes followed by a very long  almost fully absorbed GP trough 
very similar to what is found  in the observed spectrum.

\begin{figure*}
   \begin{center}
      \includegraphics[width=\textwidth,height=\textheight,keepaspectratio]{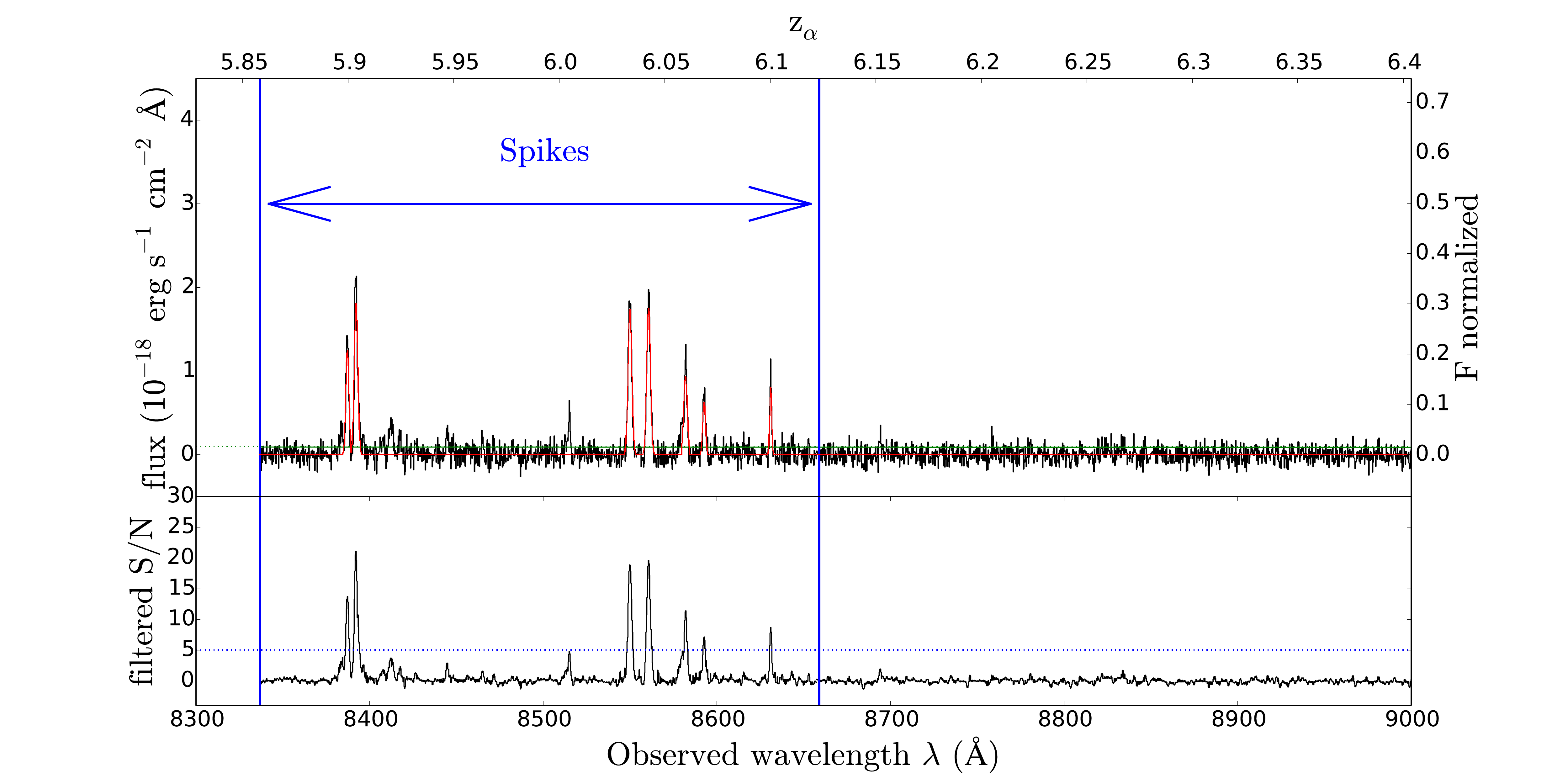}   
  \caption{Top : Zoom-in of one our  synthetic spectra of the wavelength region of  the observed spectrum of ULAS J1120+0641 showing transmission spikes.
The original smoothed plus rebinned plus noise spectrum are  shown in black. The horizontal green solid line shows the 
assumed noise level corresponding to the average noise value in the observed spectrum of  ULAS J1120+0641 as presented by  B17.
The red solid curves  corresponds to the results of Gaussian line fitting with the best $\chi^2$ values.
Bottom : the corresponding filtered signal-to-noise ratio.
}
    \label{line_fitting}
  \end{center}
 \end{figure*}

\begin{figure*}
   \begin{center}
      \includegraphics[width=\textwidth,height=\textheight,keepaspectratio]{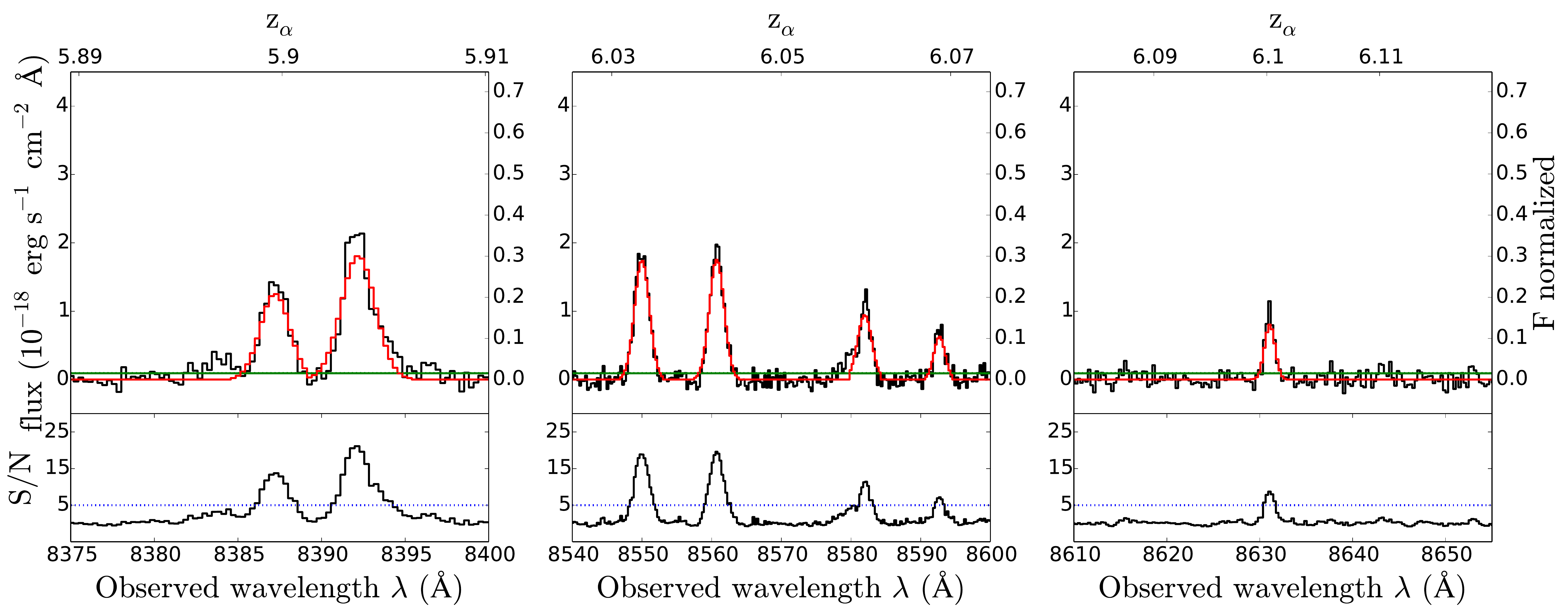}   
  \caption{Zoom-in on the different transmission spikes from Fig. \ref{line_fitting}.
}
    \label{line_fitting_zoom_individual_spikes}
  \end{center}
 \end{figure*}

\subsection{Identifying statistically significant  transmission spikes} 
\label{transmission_spikes}

In this subsection we describe our search for statistically significant transmission spikes in our mock absorption spectra which follows
closely that presented by B17 for the observed spectrum of ULAS J1120+0641. 
For the search we used a Gaussian shaped matched filter (see \citealt{2004AJ....127.1860B}) with a range of widths.
As in B17, we test different width for the Gaussian shaped matched filter with the narrowest profile having $\sigma=15 \, \mathrm{km \, s^{-1}}$
and the others having $\sigma$ increasing by factors of $\sqrt{2}$ ($\sigma=21, 30,...,120 \, \mathrm{km \, s^{-1}}$). 

For each pixel, we then record the S/N of the matched filter search and keep the most significant detections from the search with the templates with different widths. 
In Fig. \ref{line_fitting}, we show the corresponding fit  with the matched filter together with  the  mock absorption spectrum
for the part of the spectrum where transmission spikes  occur in the observed spectrum of  ULAS J1120+0641.
We also show the S/N we computed assuming a noise level equal to the value of the averaged observed noise spectrum  along the wavelength range considered. In the particular example shown, we have identified several transmission spikes similar to those in  the observed spectrum 
with a S/N$>5$  (the detection threshold  used  in B17).

Fig. \ref{line_fitting_zoom_individual_spikes} shows a zoom-in  on the seven spikes identified  in Fig. \ref{line_fitting}.
This figure can be directly compared to  Fig.~5 in B17 that shows a zoom of the seven spikes found in the observed spectrum.
The spikes in our simulated spectra and those in the observed spectrum look very  similar. 
As we will discuss in more detail later the spikes in our simulated spectra appear to  have, however,  a somewhat smaller height and   larger width 
than those in the observed spectrum.

\begin{figure}
   \begin{center}
      \includegraphics[width=\columnwidth]{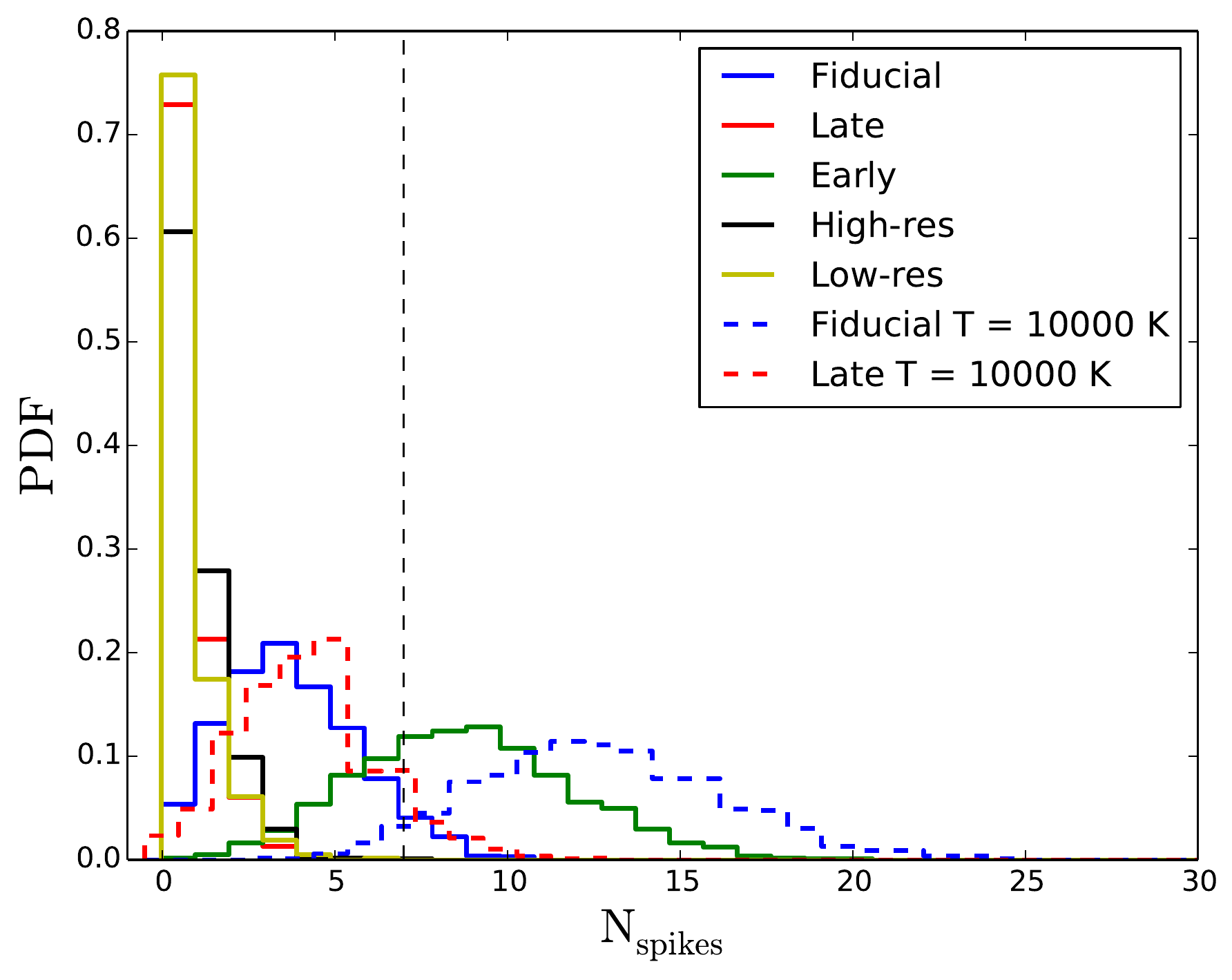}   
  \caption{The PDF of the number of identified transmission spikes obtained from a sample of 1000 line of sights covering the wavelength range 
where transmission spikes occur in the observed spectrum of ULAS J1120+0641.
The black vertical dashed line shows the value (seven transmission spikes) found in the study of B17.}
    \label{PDF_nspikes}
  \end{center}
 \end{figure}

\begin{figure}
   \begin{center}
      \includegraphics[width=\columnwidth]{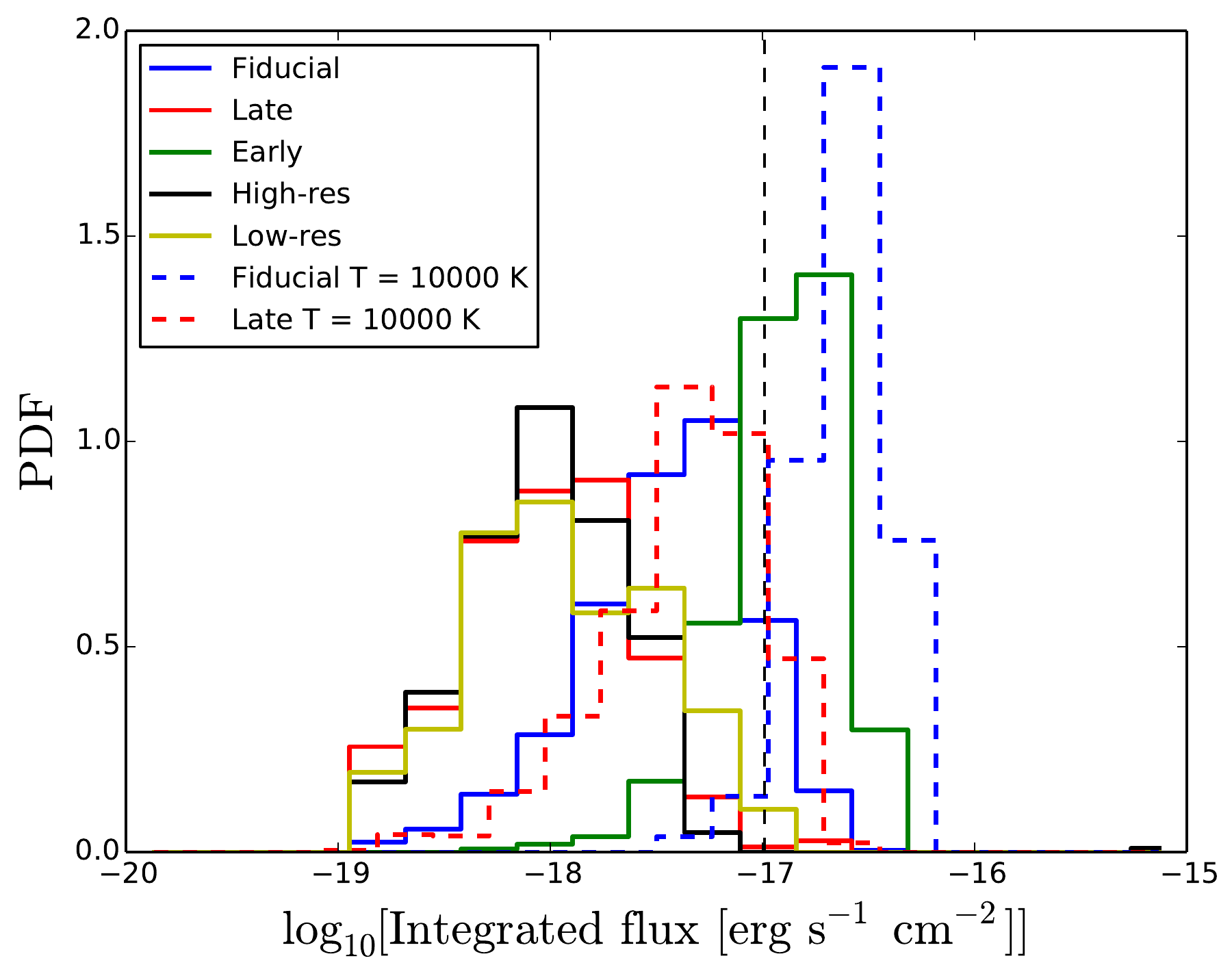}   
  \caption{The PDF of the total flux integrated over all the detected transmission spikes from a sample of  1000 line of sights through the simulations
The black vertical dashed line shows the value found in the study of B17 
corresponding to a total transmitted flux of $\mathrm{F = 1.02 \times 10^{-17} erg\,s^{-1}\,cm^{-2}}$.
}
    \label{PDF_flux}
  \end{center}
 \end{figure}

\section{Results}
\label{results}

In this section we present our main results.
First we look at the incidence rate  of transmission spikes in the different simulations and the total amount of transmitted flux. 
We then investigate whether we find a  Gunn-Peterson trough as long as in the observed spectrum.

\subsection{Incidence rate of  transmission spikes}
\label{occurence_spikes}

In Fig. \ref{PDF_nspikes}, we show the normalized probability distribution function (PDF) 
of the number of spikes in the different simulations calculated from  1000 mock spectra for random lines of sight through 
each simulation. 
The search for statistically significant transmission spikes  is performed over the wavelength range corresponding 
to the observed detection of spikes  in ULAS J1120+0641. We count a transmission spike as statistically significant 
if  at least one pixel has S/N$>5$ as in the observational study. 
The black vertical dashed line shows the value of seven transmission spikes identified by B17  in the observed spectrum of ULAS J1120+0641.

The first thing to note is that the incidence rate  of statistically significant transmission spikes is extremely  sensitive to the reionization history. 
Recall that our reionization  histories are actually all rather similar and were all chosen to be consistent with the observed overall evolution of the \lya optical depth
in observed high-redshift QSO spectra.  Interestingly, the model that appears to best match the data is the `Early' reionization history and not
our fiducial reionization history  that matches  best the evolution of $\tau_{\mathrm{eff}}$ inferred from the observed spectrum in B17 
(see Eq. \ref{teff_B17} and Fig. \ref{teff}), but this is  of course of limited significance for only one line-of-sight.  
As expected, the larger $\tau_{\mathrm{eff}}$ is in the simulation, the smaller the number of spikes detected in the skewer. 
Our fiducial  model is certainly still consistent with the observed number of spikes, but the peak of the PDF is at somewhat  lower values at  $\sim$ 2-3 spikes. It is important to note here again that we are using the temperature  from the hydro-simulations and reionization 
occurs rather early with the (homogeneous) HM2012 UV background employed in the hydro-simulations. 
The temperatures in the under-dense regions are thus rather low (about 5000K). 
As discussed by \citet{2015ApJ...813L..38D} and \citet{2017arXiv150902523D} (in prep.) temperatures in recently ionised under dense regions may be significantly hotter than this. 
For a fixed photo-ionisation rate this will  decrease the neutral hydrogen fraction and thus the \lya opacity. 
To test this we have run two further simulations where we have fixed the temperature of the gas in  ATON simulations 
of our fiducial and  late reionization histories at 10000K, independent of the density of the gas.  The resultant PDF of the number of 
spikes are shown  as the dashed blue and red curves. As expected with higher temperatures the transmissivity of the under-dense region 
responsible for the spikes increases and the PDF shifts to  larger numbers. For a temperature of 10000K 
the best match lies somewhere between the fiducial and late reionization histories while even higher temperatures 
may suggest an even later end of reionization in the line-of-sight to ULAS J1120+0641.
Finally, our lower and higher resolution simulation with a reionization history similar to that of the late reionization model show very similar results  
to the simulation at our default resolution  demonstrating that the number of significant transmission spikes  is rather insensitive to the resolution 
of the simulation.

\subsection{Total flux in transmission spikes}
\label{occurence_spikes}

\begin{figure}
   \begin{center}
      \includegraphics[width=\columnwidth]{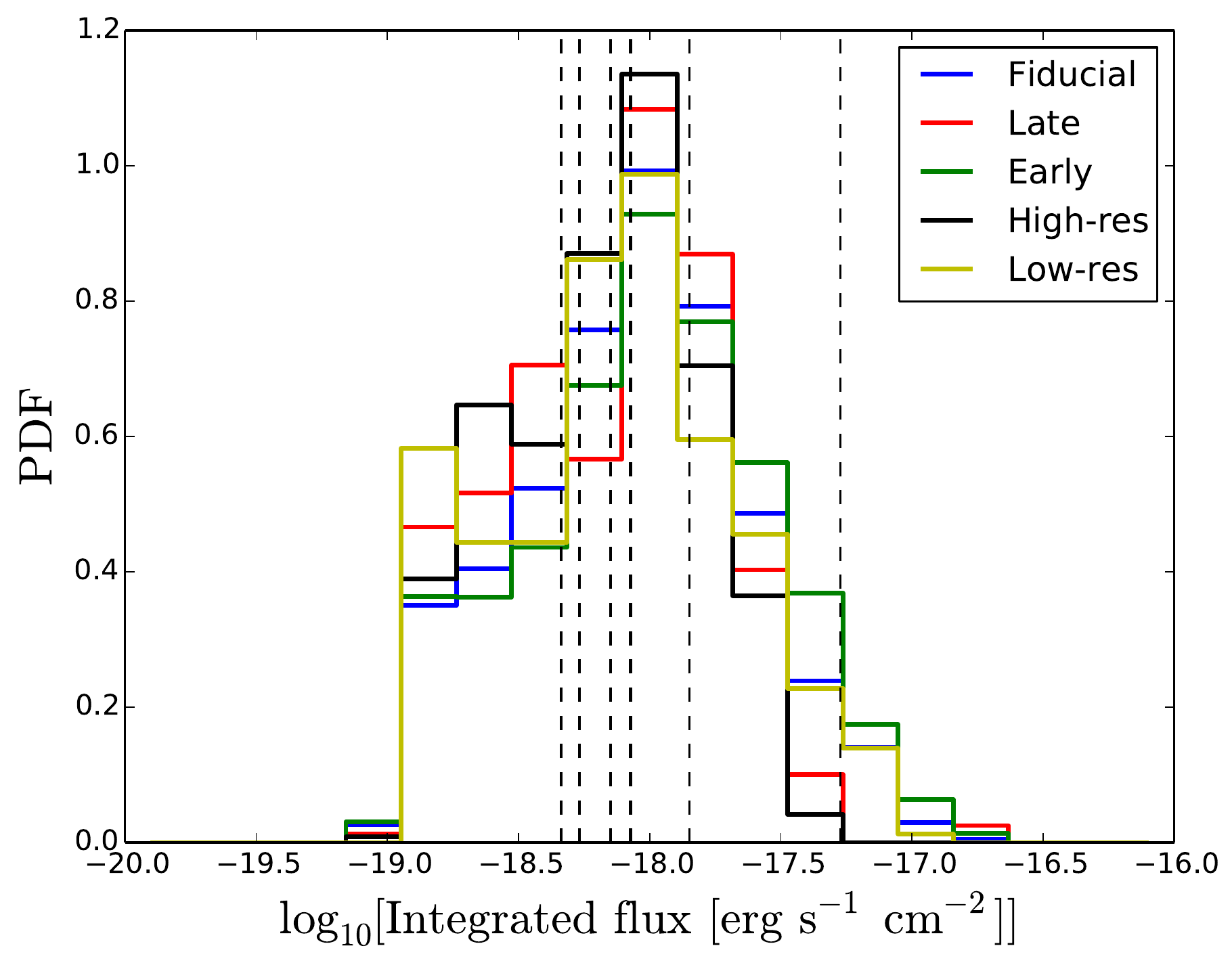}   
  \caption{The PDF of the  flux of  individual transmission spikes calculated from  1000 line-of-sights through the simulation
  of each model. The seven black vertical dashed lines show the value found in the study of B17 
corresponding to the seven spikes identified in the observed spectrum of  ULAS J1120+0641.
}
    \label{PDF_flux_individual_spikes}
  \end{center}
 \end{figure}

In Fig. \ref{PDF_flux}, we show the PDF of the total transmitted flux of all transmission spikes detected in a given spectrum 
as obtained from 1000 line of sights through the simulations with  the different reionization histories. The vertical black dashed line shows the value of $(1.02\pm0.03)\times10^{-17} \mathrm{erg \, s^{-1} \,cm^{-2}}$ reported by B17 for the observed  spectrum of ULAS J1120+0641.
As for the number of transmission spikes, we observe that the early reionization history model is the one that best matches the observed 
total flux from the real spectrum with  the low temperatures of the RAMSES simulation. 
For our fiducial reionization history the total flux is also consistent with the observed transmitted flux,  but the peak of the PDF  is located  at values 
somewhat lower than the total flux in the observed spectrum.
As shown by the dashed blue and red curves for a temperature of 10000K in the under dense regions 
the best match lies somewhere between the fiducial and late reionization histories,
similarly to what happened in case of the number of transmission spikes.

In Fig. \ref{PDF_flux_individual_spikes}, we show the PDF of the transmitted flux of  invidual transmission spikes for the different reionization 
histories. The vertical black dashed lines correspond to the transmitted flux values in the seven transmission spikes in the observed spectrum. 
All models appear to  be consistent with the values reported for  the observed spectrum. This is perhaps not  surprising as the minimum S/N for a significant detection 
corresponds to a lower limit of the flux for a transmission  spike to be counted as significant detection that depends only weakly on the width of 
the transmission spike.  Note, however, that  the observed spike with the highest transmitted flux dominates the total observed flux.

\subsection{The width  of the transmission spikes}

\begin{figure}
   \begin{center}
      \includegraphics[width=\columnwidth]{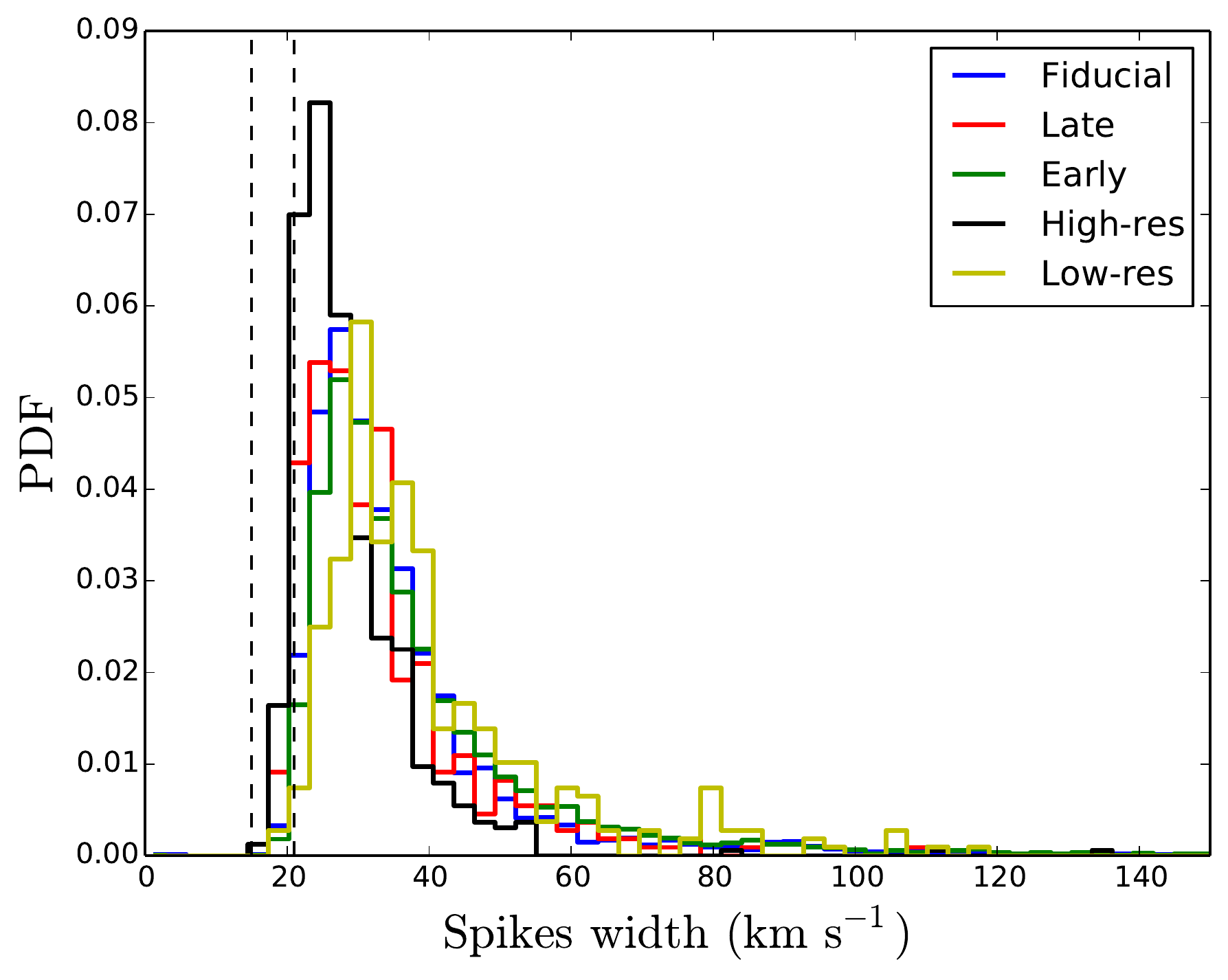}   
  \caption{The PDF of the  width of the transmission spikes as identified with a Gaussian shaped matched filter  calculated from 1000 line-of-sights
  for each model,  The two black vertical dashed lines show the value found in the study of B17 
corresponding to the seven spikes observed in the ULAS J1120+0641 quasar spectrum.
}
    \label{PDF_line_width_individual_spikes}
  \end{center}
 \end{figure}

In Fig. \ref{PDF_line_width_individual_spikes}, we show the PDF of the width of the \lya line for  transmission spikes with S/N$>$5
in terms of the $\sigma$ of the best-fit  Gaussian.  We note here that this is somewhat different to what was done by  B17 
for the observed spectrum, as they  give the width of the best fitting  Gaussian of their (rather coarse) matched filter  search while we give  
the value of $\sigma$ of the best fitting  Gaussian for each transmission spike identified.  
The values reported  for the seven spikes by  B17 are 15 $\mathrm{km\, s^{-1}}$ for six of the transmission spikes 
and 21 $\mathrm{km\, s^{-1}}$ for the remaining spike with the highest transmitted flux. We show those two values in 
Fig. \ref{PDF_line_width_individual_spikes} with the  two black vertical dashed lines. At our fiducial resolution, the spikes in our mock spectra are clearly broader 
by a factor 1.5 to 2 than the observed transmission spikes. Looking at the higher and lower resolution simulations of the late reionization history 
shows, however, that the width of the transmission spikes are not converged and decrease with increasing resolution. This is consistent  with the 
results of \citet{2009MNRAS.398L..26B} who also found that very high resolution (in their case $512^3$ for a 10 Mpc/h SPH simulation or better) 
is needed to resolve in particular the under dense regions at high redshift.  Note, however, that \citet{2009MNRAS.398L..26B} looked at 
\lya forest statistics for significantly higher photo-ionisation rates and thus mean flux levels.

\subsection{The length of totally absorbed Gunn-Peterson troughs}
\label{l_troughs}


\begin{figure}
   \begin{center}
      \includegraphics[width=\columnwidth]{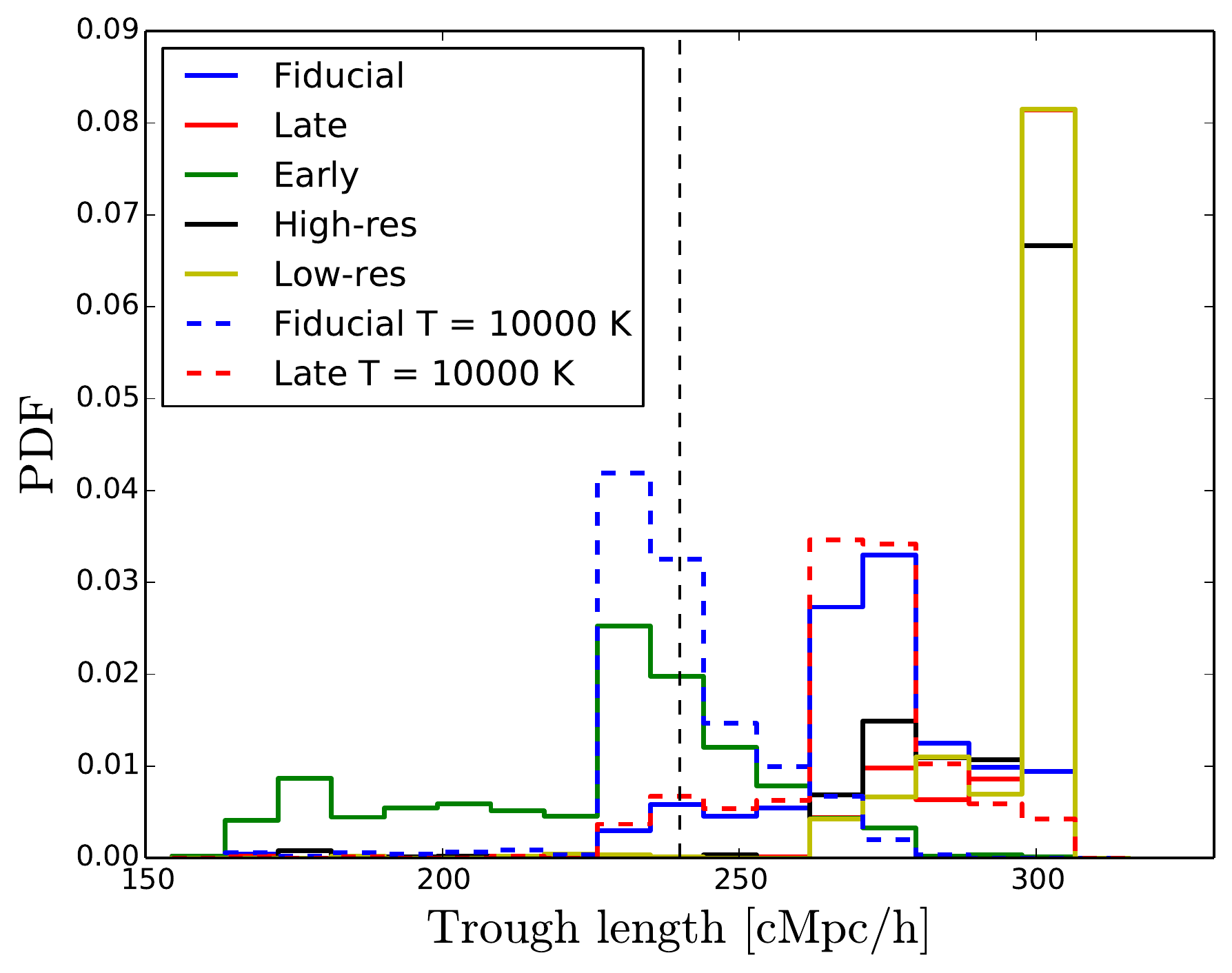}   
  \caption{The PDF of the fully absorbed GP trough length in the different models.
The values are calculated as the comoving length between the values of $z_{\mathrm{min}}$ which corresponds to highest redshift where a statistically 
significant transmission spike is identified  and $z=7.04$ 
corresponding to the edge of the proximity zone of ULAS J1120+0641. 
The black vertical dashed line shows the value found in the study of B17, $240$ cMpc/$h$.
}
    \label{PDF_length_trough}
  \end{center}
 \end{figure}

In Fig. \ref{PDF_length_trough}, we show the PDF of the length of the fully absorbed Gunn-Peterson trough for all our models.
In practice the trough lengths are calculated by finding the highest redshift of a statistically significant transmission spike in the different mock spectra.
Then we calculate the corresponding comoving length from that redshift up to a redshift $z=7.04$ that corresponds to the edge of the observed quasar proximity zone.

Again, we note  that it is the simulated spectra with the early reionization history  that best match the observed trough length
for the low temperatures in under-dense regions of the RAMSES simulation.
The model with the fiducial reionization  history is also producing  troughs  with the observed lengths  although the PDF 
of the trough length peaks at somewhat  larger values. This is consistent with our finding for the 
incidence rate  of statistically significant transmission spike at the lower end of the redshift range probed by the 
\lya forest in the observed spectrum  of ULAS J1120+0641. The blue and red dashed curves show the 
PDF of the absorption trough length for the 10000K case for the fiducial and late reionization models. 
Higher temperatures in the under-dense regions  lead to a higher number of spikes detected with some of them found at higher redshift.
This results in lower length ot GP trough in the spectra.

\section{Discussion}
\label{discussion}

\subsection{The \lya forest of $z>6.5$ quasars to constrain the reionization history}
\label{QSO_probe_reion}

In the last sections we have shown that matching the absorption properties  of the \lya forest in ULAS J1120+0641 in radiative 
transfer simulation is very sensitive to the reionization histories assumed in the simulations.  It also depends sensitively 
on the temperature of the IGM in under-dense regions. 
The use of narrow transmission  
spikes only detectable with high-SN high resolution spectra pushes the redshift limit  where information about  the ionisation 
state of the hydrogen in IGM can be extracted significantly higher. This suggests that obtaining  more 
high-SN spectra of $z>6.5$ QSOs should  allow to put  strong constraints  on the timing of the overlap of HII bubbles in many lines-of-sight.
Currently  ULAS J1120+0641 is the only QSO  with an 30h high-resolution spectrum.
There are already ten quasars found at $z\geq6.5$. Four were discovered in near-infrared surveys 
(the quasar at z=7.085 studied in the current paper  was first reported in \citealt{2011Natur.474..616M} and three other quasars were reported in \citealt{2013ApJ...779...24V}). 
Three new quasars (\citealt{2015ApJ...801L..11V}) were discovered recently from the 3 $\pi$ Panoramic Survey Telescope 
and Rapid Response System (\citealt{2010SPIE.7733E..0EK}) (Pan-STARRS1 or PS1). 
Two new faint quasars have been discovered by the Subaru High-z Exploration of Low-Luminosity Quasars (SHELLQs) survey
(\citealt{2016ApJ...828...26M},\citealt{2017arXiv170405854M}) and one by the Dark Energy Energy survey (DES; \citealt{2017MNRAS.468.4702R}). 
Moreover, a new high-redshift quasar, PSOJ006.1240+39.2219, at z=6.61 has recently been found (\citealt{2017MNRAS.466.4568T},\citealt{2017arXiv170605785K}). 
Re-observing the nine additional  quasars at $z\geq6.5$ with high SN can thus be expected  to extend 
our ability to  use the  \lya forest as probe of the tail-end of reionization to significantly higher redshift than
previously possible.

\subsection{Caveats of our modelling}
\label{caveats}

The models presented here all assume that  reionization is driven by ``star-forming'' galaxies. 
However, motivated by the large observed opacity fluctuations at rather larger 
scales (\citealt{2015MNRAS.447.3402B}) some authors including us have  recently 
revisited the old idea  that QSOs may  contribute significantly to the ionising emissivity 
at high redshift  (\citealt{2015AA...578A..83G} and \citealt{2015ApJ...813L...8M}, but see also  \citealt{2017arXiv170405996} and \citealt{2017arXiv170407750P} for 
alternative views on  the space density of high-redshift AGN found by \citealt{2015AA...578A..83G}).
As discussed in detail  by \citet{2015MNRAS.453.2943C} and \citet{2017MNRAS.465.3429C},
a significant contribution of QSOs to the ionising emissivity at $z\ga 5.5$ could  lead to large fluctuations of the photo-ionisation 
rate well after the percolation of HII bubbles  and could explain the broad distribution  of the effective optical depth  $\mathrm{\tau_{eff}}$ measured in 
50 cMpc/$h$ chunks by \citet{2015MNRAS.447.3402B}. If necessary, modelling a possible  QSO contribution 
to the ionising emissivity will obviously further complicate the interpretation of the observed transmission spikes.
Note, however,  that there is currently  no evidence for  the presence of QSOs close
to the line of sight of ULAS J1120+0641 even though   QSOs contributing to the ionising emissivity in the line-of-sight 
to ULAS J1120+0641  could have obviously switched off or may not point towards us.

Two alternative  explanations have been put forward for the observed large opacity fluctuations on large scales:
spatial fluctuations of the temperature-density relation of the IGM  (\citealt{2015ApJ...813L..38D},  but see \citealt{2017arXiv150902523D} (in prep.))
and  large mean-free path fluctuations that could arise when the  mean free path of ionising photons is still shorter than or comparable 
to the mean distance between ionising sources  (see \citealt{2016MNRAS.460.1328D} and \citealt{2016arXiv161102711D}).
Modelling this properly is rather difficult and beyond the scope of this paper but may also further complicate the interpretation of the observed transmission spikes.

\section{Conclusions}
\label{conclusion}

We have presented here the result of our analysis of  mock \lya absorption spectra that aim  to reproduce the observed high S/N  
high-resolution spectrum of the  $z=7.084$ QSO  ULAS J1120+0641 recently presented  by B17. 
The radiative transfer simulation used to construct the mock absorption spectra   had been previously carefully calibrated to match the main  
observational constraints  from \lya forest data   such as the redshift evolution of the effective optical depth, the neutral hydrogen fraction,
the inferred  average  hydrogen photoionization rate and the Thompson optical depth measured from the CMB. 

Our main results are as follows.
\begin{itemize}
  \item{The occurrence of narrow transmission spikes from highly under dense regions and the onset of a fully absorbed GP trough 
          in our simulations is caused by the rather rapid rise of the photionization rate following the percolation of HII regions and 
          is very  sensitive to the exact timing of the overlap of HII bubbles as well as  the temperature of the IGM at low densities.}       
 \item{Using our well calibrated simulations with a narrow range of three reionization histories (early, fiducial and late), we were able to reproduce the 
occurrence of seven transmission spikes, the observed integrated transmitted flux and the length of the totally absorbed 
GP trough in the observed spectrum of ULAS J1120+0641 very well.}
\item{Reionization histories with a later overlap of HII region require the under-dense regions to be hotter to be consistent 
         with the observed spectrum.}  
\item{The width of the transmission spikes in our simulated spectra is not fully converged and is about a factor 1.5-2 larger 
          in spectra produced from our simulations with our fiducial resolution ($\mathrm{512^3}$ grid cell, 20 cMpc/h box size)}.
\end{itemize}

Our analysis shows that identification of transmission spikes in high SN spectra can push the constraints on the ionisation state of the IGM 
to redshifts as large as $z=6.1$ and possibly higher, where the average \lya optical depth is already very large. 
This  is significantly higher than was previously possible and constraints on the ionisation state of the IGM from \lya forest data 
at these redshifts were so far limited  to the difficult analysis of the near-zones and possible red damping wings in the QSO absorption 
spectra. Extending the search for and the modelling 
of transmission spikes at $z>5.8$ in (still to be obtained) high-SN absorption spectra of a larger sample of QSOs should provide 
important insight into the exact timing of the overlap of HII bubbles at the tail-end of reionization.

\section*{Acknowledgments}

We thank James Bolton and George Becker for comments on the manuscript. 
This work was supported by the ERC Advanced Grant 320596 ``The Emergence of Structure during the epoch of Reionization". 
The RAMSES simulation were performed utilizing the supercomputer COSMOS Shared Memory system at DAMTP, University of Cambridge operated on behalf of the STFC DiRAC HPC Facility. 
This equipment is funded by BIS National E-infrastructure capital grant ST/J005673/1 and STFC grants ST/H008586/1, ST/K00333X/1.
The ATON radiative transfer simulation in this work were performed using the Wilkes GPU cluster at the University 
of Cambridge High Performance Computing Service (http://www.hpc.cam.ac.uk/), 
provided by Dell Inc., NVIDIA and Mellanox, and part funded by STFC with industrial sponsorship from Rolls Royce and Mitsubishi Heavy Industries.

\bibliographystyle{mnras}
\bibliography{biblio}

\appendix

\section{The remaining flux below detection limit}

\begin{figure}
   \begin{center}
      \includegraphics[width=\columnwidth]{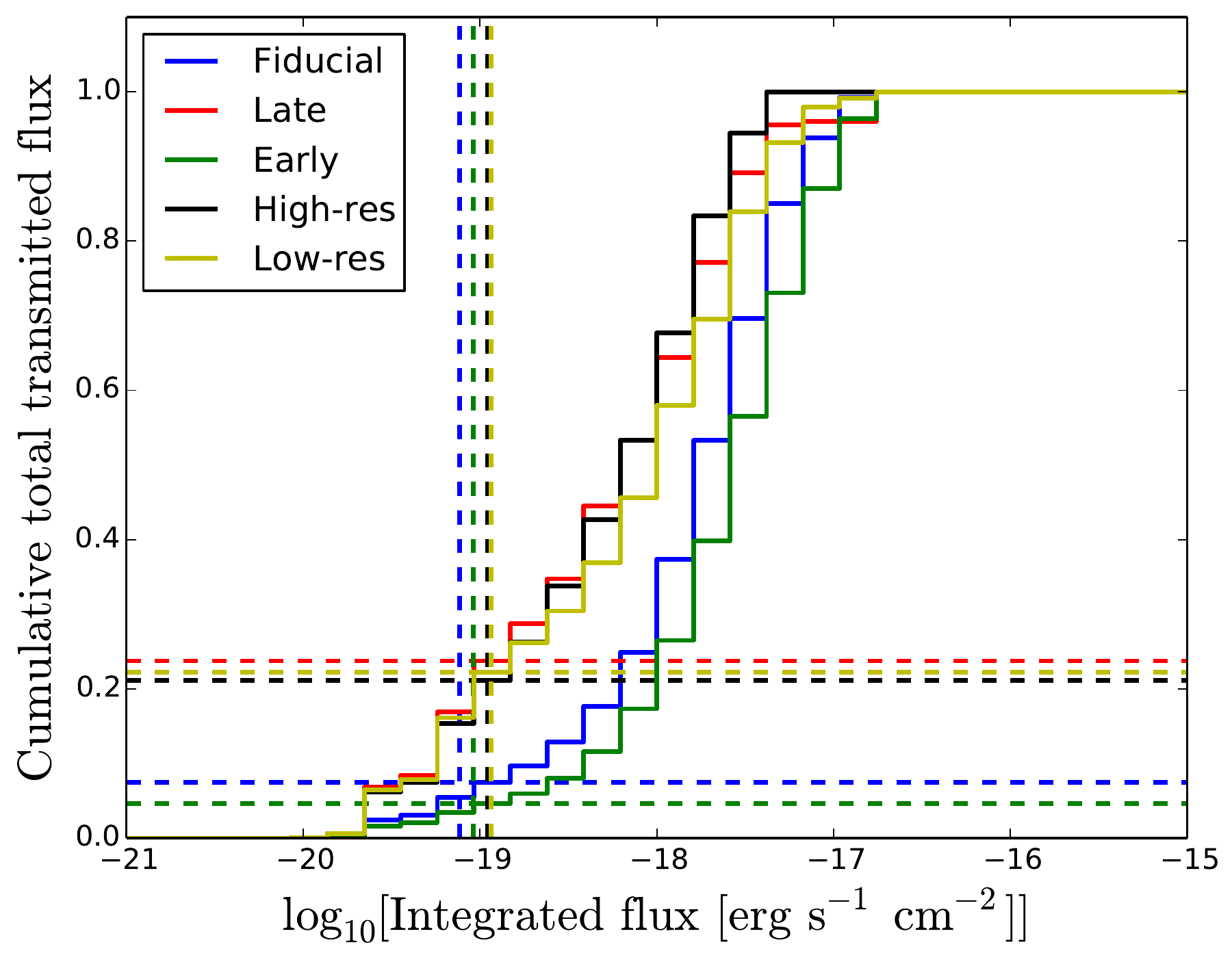}   
  \caption{The cumulative transmitted flux as a function of flux in individual transmission spikes in the different models, if  we assume a signal to noise ration S/N=1 in order to estimate the flux in transmission that is beyond the current detection limit of  the observations by B17.
The vertical coloured dashed lines show the detection limit of the flux in individual spikes assuming a signal to noise ration S/N=5 
(the smallest values found in each simulation in Fig. \ref{PDF_flux_individual_spikes}) as in the observations.
The corresponding horizontal coloured dashed lines show the percentage of total flux that would be missed by current observations in the different models.
}
    \label{PDF_cumul_flux}
  \end{center}
 \end{figure}

In Fig. \ref{PDF_cumul_flux}, we show  the cumulative transmitted flux for  all the different models as a function of the flux in individual transmitted spikes. In order to estimate the amount of flux  beyond the detection limit of the search performed by B17, we use a signal to noise ratio S/N=1.
To this end we have rerun a  Gaussian matched filter search with this value of the S/N. 
The vertical dashed lines in Fig. \ref{PDF_cumul_flux} shows the flux detection limit of the spikes detected with S/N=5 as in the observations 
(corresponding to the smallest value of flux in individual spikes found in Fig. \ref{PDF_flux_individual_spikes}).
The corresponding horizontal dashed lines show the percentage of the total transmitted flux that is  beyond the detection limit 
of Barnett et al. (2017).

About 5 and 7 \% of the total transmitted flux are missing  for the  `Fiducial' and `Early' reionization histories, respectively.
Larger values of the order of 20 \% are found for simulations with the  `Late' reionization history and for the , `Low-res' and `High-res' simulations.
These larger  values are  not surprising as spikes with large values of transmitted flux are much rarer in these simulations because reionization ends to late to be consistent with the observations. 
Therefore,  small spikes beyond the detection limit impact  the total transmitted flux more significantly in these simulations. 
Since these models are, however,  not fitting the observed number of transmission spikes, we can conclude that our simulations suggest that the the majority (about 95\%) 
of the total transmitted flux in ULAS J1120+0641  has been detected by  Barnett et al. (2017).

\end{document}